\documentstyle[12pt]{article}

\begin{document}
\setcounter{page}{0}
\begin{flushright}
{CERN--TH.7200/94}\\
{BI--TP 94/04}
\end{flushright}
\medskip
\begin{center}
{\bf MEASURING THE HADRONIZATION LENGTH IN $e^+ e^- \rightarrow K^0 {\bar{K}}^0
\gamma$  }
\end{center}
\vskip1cm
\begin{center}
{\bf Dmitri Kharzeev\footnote{On leave from Moscow State University, Moscow,
Russia}$^{(a)}$
 and Andrei Leonidov\footnote{Alexander von Humboldt Fellow}$^{(b)}$}
\end{center}
\begin{center}
{\it (a) Theory Division\\CERN\\ CH-1211, Geneva 23\\ Switzerland}
\end{center}
\medskip
\begin{center}
{\it (b) Physics Department\\ University of Bielefeld, 33615
Bielefeld, Germany\\ and \\Theoretical Physics Department\\ P.N. Lebedev
Physics Institute, Leninsky pr. 53, 117924 Moscow, Russia}\\
\end{center}
\medskip
\begin{abstract}
We propose to study the space--time picture of hadronization by looking at the
angular
distribution of photons emitted in the process $e^+ e^- \rightarrow
K^0 {\bar{K}}^0 \gamma$ at DA${\Phi}$NE energies. It is shown that the angular
distributions are quite sensitive to the hadronization length.
\end{abstract}
\vskip3cm
CERN--TH.7200/94\\
March 1994
\newpage

The phenomenon of the confinement of quarks and gluons is one of
the most exciting and at the same time mysterious features of particle
physics. Its analysis goes in two main
directions. The first is the proof of its existence in the framework
of local quantum field theory of strong interactions, Quantum
Chromodynamics. The second one is more phenomenological and deals with
the experimental consequences of confinement. Since the property of
confinement means that free--colour--charge states do not exist and since at
the same time the early stage of many processes is well described by
weakly interacting quarks and gluons, the natural object to study is the
process of conversion of the quark-gluon degrees
of freedom into hadronic ones (hadronization).
The space-time picture of it is of particular interest. It is clear that
finding
a signal coming from the quark-gluon stage of the process
and significantly dependent on the hadronization scale could provide a unique
information
 on the hadronization length.  One of the most
promising signals is the electromagnetic radiation in the processes
where the electrically neutral hadrons (hadron jets) are formed [1].
In this case the dominating signal will come from the hard
stage (i.e. the photons will be emitted by quarks before they become
neutral hadrons) and studying the properties of this radiation one
learns about the hadronization itself. In this Letter we propose to study a
specific process of this type, which could give an especially clean
signal and thus provide direct experimental information on
quark confinement and hadronization.

The particular process we are interested in is the
electron-positron annihilation into two neutral kaons and a gamma quantum
$e^{+}e^{-} \rightarrow K^{0} {\bar{K}}^{0} \gamma$ (see Fig. 1) in the
energy range
below the $K {\bar{K}}^{0*}$ production threshold
and above the $\phi$ resonance, i.e. in the energy
interval of $1025 \mbox { MeV } < Q < 1300 \mbox { MeV },$ where $Q$ is
the energy of
$e^{+}e^{-}$ annihilation in the centre--of--mass system. Why is this
particular process so convenient in studying the
space--time picture of the hadronization? To answer this question
we first discuss its physical picture.
The strange quark-antiquark pair produced initially is only {\it{later}}
converted into a pair of neutral kaons. Let us note that
at these energies the velocity of the produced strange quark and antiquark is
high enough (of the order of 0.76$c$) for them to be described,
due to asymptotic freedom, as initially free particles.
 If the neutral kaons are produced in their ground
state, the only source of final--state radiation is the quark or antiquark
produced initially.
Since in the considered energy range
we are far from any resonance, there should be no
background photons from their decays. Thus apart from the bremsstrahlung
radiation of the
initial electron-positron beams there seems to be no background to the
radiation of the quark (antiquark) emitted {\it{before}} they
hadronize and become neutral kaons. Let us stress that this picture is
an assumption, and that whether this understanding of
hadronization in the discussed process is correct  should be
revealed by the experiment. The
transition from the completely coherent processes at
very low energies, where the excited degrees of freedom are hadronic ones
from the very beginning, to the regime where the kinematical structure of
the event is completely determined by that of
the initial hard quarks and gluons at high energies (jets) is completely
non--trivial and deserves careful analysis.
 The way to
understand the problem is thus to compare the model predictions  with
the experimental data. If we believe that, at the initial stage of the
process, we are dealing with the charged quark and antiquark which then become
neutral kaons, the fact that the radiation is emitted within some
typical length manifests itself in the change of its characteristics
first discussed by Frautschi and Krzywicki [2] and by Dremin [5], and then in
a more usual
field theory framework in [3,1]. A concise review of the damped currents
radiation and of other phenomena connected with the radiation in
the medium as applied to high energy collisions can be found in [4].
Recently, the importance of accounting for the finite scale of colour
neutralization in relativistic nuclear collisions was outlined in [6].

Let us now turn to a quantitative analysis of the process described above.
In order to calculate the properties of the final
state photon radiation in the process $e^{+} e^{-} \rightarrow K^{0}
{\bar{K}}^{0} \gamma,$  taking into account the final hadronization
length, we shall use the model quark propagator where the finite
radiation length corresponds to the imaginary part in the denominator [3,1]
\begin{equation}
G_{q} (p)={i ({\hat{p}}+m) \over p^2-m^2+i m \Gamma}
\end{equation}
describing the propagation of an unstable quark with a lifetime of
order $1/\Gamma$. Let us stress that this is only one of the possible
ways of considering the finite decolorization (and in the present
case also the electrical charge neutralization) length (see [1-4]).

It is easy to acquire the conviction that in the CMS energy range under
consideration only soft photons can be emitted. This allows the use of a
well-known eikonal formula for the soft photon radiation
\begin{equation}
w(p_q,p_{{\bar {q}}},k)=- \left({p_q \over (p_q k)}-{p_{{\bar{q}}} \over
(p_{{\bar{q}}} k)} \right)^2,
\end{equation}
where $p_q$, $p_{{\bar{q}}}$ and $k$ are the four-momenta of the quark,
antiquark and photon respectively.
When
using a propagator (1), we get instead of (2)
\newpage
\begin{eqnarray}
\omega(x_1,x_2) & = &
{4 \over Q^2} \left( {(x_1+x_2-1-2\mu^2)((1-x_1)
(1-x_2)+\mu^2
\gamma^2_q) \over
((1-x_1)^2+\mu^2 \gamma_q^2)((1-x_2)^2+\mu^2 \gamma_q^2)} \right. \nonumber \\
 &   & \left. -{\mu^2((1-x_1)^2+(1-x_2)^2+2 \mu^2 \gamma_q^2) \over
((1-x_1)^2+\mu^2
\gamma_q^2)((1-x_2)^2+\mu^2 \gamma_q^2)} \right),
\end{eqnarray}
where $\mu=m/Q$, $\gamma_q=\Gamma/Q$
and we have introduced the standard variables $x_{1,2}=
2 E_{q,{\bar{q}}}/Q$. In the following we shall use the value of the
strange quark mass $m=0.5 \mbox { GeV. }$

A useful observable distribution that we can obtain from (3) is a
double differential distribution in the
energy fraction $M$ carried by the less energetic kaon (antikaon)
and the angle  $\theta_{M\gamma}$
between this kaon and the
emitted photon. From Eq. (3) we get for this
distribution:
\begin{eqnarray}
W(M,\mbox {cos} \theta_{M \gamma})&=&
 {16 M^3 (1-M) \over (M+{\mbox {cos}} \theta_{M \gamma}
 (M^2-4 \mu^2)^{1/2})^{2}} \nonumber \\ &&
 \Theta \left( \mbox {cos} \theta_{M \gamma}+{1-M \over (M^2-4
\mu^2)^{1/2}} \right) \omega (M,x_2(M,\mbox {cos} \theta_{M \gamma})),
\end{eqnarray}
where $\Theta$ is a step function and
\begin{equation}
x_2(M,\mbox {cos} \theta_{M \gamma})={2-(1-M)(M-\mbox {cos} \theta_{M
\gamma} (M^2-4\mu^2)^{1/2}) \over 2-M+\mbox {cos} \theta_{M \gamma}
(M^2-4 \mu^2)^{1/2}}.
\end{equation}
In Fig. 2 we show a distribution in $\mbox {cos} \theta_{M \gamma}$
obtained at $Q=1.3 \mbox { GeV, }$ $\Gamma=0.23 \mbox { GeV }$ and $M=0.9,$
and at $\Gamma=0$. We see that the account for the decolorization
length leads to an angular distribution which is noticeably flatter and
less in amplitude than the one corresponding to the infinite
hadronization length. From Fig. 2 one also observes a
reduction of the total radiation cross section. In Fig. 3 we also present a
two-dimensional
distribution
$W(M,\mbox {cos} \theta_{M
\gamma})$ at $\Gamma=0.23 \mbox { GeV. }$

We can conclude that the angular distribution of the
$\gamma$-quanta in the process $e^{+}e^{-} \rightarrow K^{0}
{\bar{K}}^{0} \gamma$ is indeed very sensitive to the (hadronization)
screening length,
thus giving a hope for the experimental observation of this phenomenon at
DA${\Phi}$NE
energies.

Let us summarize the results of the paper.
We propose to study the space-time dynamics of hadronization by studying
the process $e^+ e^- \rightarrow K^0 {\bar{K}}^0 \gamma$. By using
the simplest model description of the finite length at which the
radiation of the charged quark (antiquark) is possible, we obtain the
angular distribution of the emitted photon with respect to a direction
defined by a less energetic kaon. We have shown that the angular
distribution is  sensitive to the scale at which the hadronization
takes place, thus opening a possibility of  the experimental
study of confinement at DA$\Phi$NE energies.

Finally let us briefly mention what roads for future research
 we see.

1. The most obvious improvement should be a calculation including the
radiation of the initial beams and corresponding interference, thus providing
a quantitative basis for
the separation of the interesting contribution.
However the bremsstrahlung of the initial electron and positron beams
is confined within a narrow collinear cone; this leaves
a possibility to perform the proposed measurement at relatively large (with
respect to the colliding beams) angles.

2. In the above consideration we have neglected the influence of the
vacuum quark (antiquark) on the momenta of the initially produced
antiquark (quark) in the process of $K^0$--meson formation. One
can hope that for the considered process this influence is not big.
However, this question definitely deserves further study.

3. One of the interesting possibilities arising in the described
context is a quantum interference between the case where the strange
pair is produced directly by $e^+ e^-$ and the case where it is taken
from the vacuum, leading in both cases to the same kaons. As the
radiation will in this case come from the quarks having different mass,
one could  observe a mixture of two contributions to the angular
distributions.

4. The same approach can easily be applied in studying the radiation of
similar currents on $B$-- and $c \tau$--factories.

5. A situation analogous to the one described in the paper takes place
in the process $e^+ e^- \rightarrow N {\bar{N}} \gamma$, but in the case
of neutrons we are looking at the vacuum production of the diquarks (or
possibly the small-size diquark configurations directly in the
electron-positron annihilation).

\begin{center}
{\it{Acknowledgements}}
\end{center}

D.K. is indebted to Prof. H. Satz for many stimulating discussions on
colour neutralization and helpful comments on the manuscript.

A.L. is grateful to Prof. H. Satz for the kind hospitality of the University
of Bielefeld, acknowledges a partial financial support by the Russian
Fund for Fundamental Research, Grant 93-02-3815, and thanks M. Bilenky and
S. Mikhailov for useful discussions.

\newpage

\begin{center}
{\it\bf{References}}
\end{center}

1. I.M. Dremin and A.V. Leonidov, {\it{Mod.Phys.Lett.}} {\bf{A1}} (1986) 51;

2. S. Frautschi and A. Krzywicki, {\it{Z.Phys.}} {\bf{C1}} (1979) 43;

3. A.V. Leonidov, {\it{Lebedev Inst. Repts.}} {\bf{5}} (1985) 59;

4. I.M. Dremin, {\it{Sov.J.Part.Nuclei}} {\bf{18}} (1987) 31;

5. I.M. Dremin, {\it{JETP Lett.}} {\bf{34}} (1981) 594;

6. D. Kharzeev and H. Satz, {\it{Z.Phys.}} {\bf{C60}} (1993) 389.

\vskip2cm
\begin{center}
{\it\bf {Figure Captions}}
\end{center}

Fig. 1 The radiation of the photon by the strange quark in the process
$e^+ e^- \rightarrow K^0 {\bar{K}}^0 \gamma$.
\vskip0.3cm
Fig. 2 The distributions $W(M,\mbox {cos}\theta_{M \gamma})$ in $\mbox
{cos}\theta_{M \gamma}$ at $Q=1.3 \mbox { GeV, }$ $M=0.9$.

\hskip2cm The solid line
corresponds to $\Gamma = 0.23 \mbox { GeV, }$ the dotted line to
$\Gamma=0$.
\vskip0.3cm
Fig. 3 The  two-dimensional distribution $W(M,\mbox {cos} \theta_{M
\gamma})$ at $\Gamma=0.23 \mbox { GeV. }$

\end{document}